# Enhanced Directional Quantum Emission by Tunable Topological Doubly-Resonant Cavities

CHENMIN XU, CHONG SHENG, SHINING ZHU, AND HUI LIU[*]

[1]National Laboratory of Solid State Microstructures, School of Physics, Collaborative Innovation Center of Advanced Microstructures, Nanjing University, Nanjing 210093, China

*liuhui@nju.edu.cn

**Abstract:** How to utilize topological microcavities to control quantum emission is one of the ongoing research topics in the optical community. In this work, we investigate the emission of quantum emitters in doubly-resonant topological Tamm microcavity, which can simultaneously achieve dual resonances at two arbitrary wavelengths according to the needs of practical application. To achieve the enhancement of quantum emission in such cavities, we have exploited the tunable doubly-resonant modes, in which one of resonant modes corresponds to the pump laser wavelength and the other one is located at the emission wavelength of quantum emitters. Both theoretical and experimental results demonstrate that the pump excitation and emission efficiencies of quantum emitters are greatly enhanced. The main physical mechanism can be explained by the doubly-resonant cavity temporal coupled-mode theory. Furthermore, we observe the faster emission rate and the higher efficiency of unidirectional quantum emission, which have promising applications in optical detection, sensing, filtering, and light-emitting devices.

Since the discovery of the quantum Hall effect, various topological states have been widely studied[1-3], such as topological insulators[4, 5] and topological semimetals[6, 7]. In addition to condensed electron systems, they have been found in various other physical systems, including cold atomic lattices[8, 9], photonic crystals[10-13], phonon crystals[14], coupled resonator arrays[15, 16], superconfigurable materials[17, 18], etc. The most important characteristic of the topology is the edge state protected by topology[19-22]. Such topological edge state can form topological resonant cavities in optical systems, which are utilized to enhance light-matter interactions and change the emission properties of materials, realizing new types of topological lasers[23-26] and quantum emission sources[27, 28]. Therefore, the use of topological microcavities to manipulate quantum emission has attracted considerable attention[29-32]. But so far, most of these topological microcavities are single resonant cavities for either the pump laser wavelength or the emission wavelength of quantum emitters (QEs)[33-39]. Intriguingly, the stronger emission of QEs can be achieved by placing them in the doubly-resonant cavity to simultaneously resonate for pump laser and emission wavelengths[40], which has been demonstrated in several works used micro-nano photonic structures, including doubly-resonant nano-antennas[41, 42] and coupled optical resonant cavities[43-45]. Although these structures can achieve resonance at two specific wavelengths, the pump laser and emission wavelengths cannot be tuned freely according to the needs of practical applications. Besides, since these doubly-resonant micro-nano structures are not topologically protected, the emission of QEs is susceptible to various defects and the external environment. Therefore, the design of doubly-resonant microcavities with topological property that can resonate at arbitrary pump laser and emission wavelengths has good practical applications. Unfortunately, to the best of my knowledge, very few works so far have reported on the doubly-resonant topological microcavity with both tunable resonance of pump laser and emission wavelengths.

Among the topological photonic systems that have been reported to data, one-dimensional photonic crystal structures, in which the unit cell is central mirror-symmetric, are simple and easy-to-implement. They can be directly excited by both TM- and TE- polarized light from free space in the arbitrary direction without the complex exciting condition for momentum matching, such as with the aid of prisms or diffraction gratings[46]. The Zak phase of the photonic energy band is either integer 0 or $\pi$, and the surface reflection phase is positive or negative[47]. When two photonic crystals with opposite surface reflection phase are spliced together, a topologically protected interface state can be obtained[48]. Likewise, the topological Tamm state can be achieved if a metallic material with negative reflection phase and a photonic crystal with positive reflection phase are spliced together[49]. The topological Tamm state has tunable resonant wavelength compared to the normal Tamm state. That is, the topological Tamm state with arbitrary resonant wavelength can be achieved by maintaining the mirror symmetry of the photonic crystal while telescoping its period. So, by utilizing the tunable topological Tamm state, can we achieve resonance at arbitrary pump laser wavelength and emission wavelength of QEs?

In this work, we first propose a method to enhance the emission of QEs using a special topological photonic system; i.e., the asymmetric doubly-resonant topological Tamm microcavity. Such a structure has a remarkable feature with two freely tunable resonant modes to simultaneously achieve dual resonances for arbitrary pump laser and emission wavelengths. We investigate the photoluminescence (PL) intensity under three different cases. One case is that the pump laser wavelength locates at the resonant wavelength, whereas the emission wavelength of QEs is away from the resonant modes. Another case is the opposite condition that the emission wavelength of QEs is at the resonant mode, while the pump laser wavelength is not. In the last case, both the pump laser wavelength and emission wavelength of QEs are at the resonant modes of our structure, which largely enhances the emission of QEs.

The structure of our design is illustrated in Fig.1(a), which comprises of two type of photonic crystal, silver and polymethylmethacrylate (PMMA) mixed with quantum emitters. The unit cells of photonic crystals are both composed of *A/B/A* layers that are mirror-symmetric, as marked by the red dashed lines in Fig.1(a). For one type of photonic crystal named by Phc1, layer A is Ta$_2$O$_5$ with a thickness of $a_1 = 33nm$ and layer B is SiO$_2$ with a thickness of $b_1 = 82nm$. The period is $\Lambda_1 = 2a_1 + b_1 = 148nm$. For the other type of photonic crystal named by Phc2, the corresponding layer of A and B are respectively $a_2 = 40nm$ and $b_2 = 105nm$, and the period is $\Lambda_2 = 2a_2 + b_2 = 185nm$. Their Zak phases of the photonic energy band are integer 0 or $\pi$. And, the reflection phases of their first-order photonic band gap are positive numbers, satisfying the condition $Z_{Phc} + Z_{Ag} = 0$, which means that the surface impedance of the photonic crystal inverse match to the silver film. Under this condition, the topological Tamm state exists at the interface between the metallic Ag and the photonic crystal. The refractive indexes of Ta$_2$O$_5$ and SiO$_2$ are from experimental measurements. In the working frequency range, the dispersion is negligible and the refractive indexes of Ta$_2$O$_5$ and SiO$_2$ are respectively around 2.14 and 1.48. The thickness and the refractive indexe of PMMA are respectively $150nm$ and 1.62. For the silver layer, the thickness is 70nm, and the complex permittivity is described as Drude model with reference characteristic parameters: $\varepsilon_m = \varepsilon_m^{'} + i\varepsilon_m^{''} = 5 - \omega_p^2/(\omega^2 + i\omega\omega_\tau)$, the plasma frequency $\omega_p = 9.1eV$ and the relaxation rate $\omega_\tau = 0.021eV$.

To achieve doubly-resonant microcavity, we first realize single topological Tamm state which is comprised of a single photonic crystal and a silver film. Fig.1(b) and Fig.1(c) correspondingly depict the topological Tamm state under the case of Phc1 and Phc2. When the polarized light is incident, there are peaks in the transmission spectra based on transfer matrix method, whose wavelengths are $\lambda_1 = 500nm$ for Phc1 and $\lambda_2 = 610nm$ for Phc2. The inset in Fig.1(b) and Fig.1(c) clearly exhibit that an optical localization effect occurs at the interface between the metal and the dielectric plane, where the electric field intensity reaches a maximum. Intriguingly, when we combine such two single Tamm cavities together to form a doubly-resonant microcavity consisting of *Phc1/PMMA/Ag/Phc2*, two peaks appear in the transmission spectrum within the Bragg band gap as shown in Fig.1(d). One of the peaks corresponds to topological Tamm state1 (TTS1) and the other to topological Tamm state2 (TTS2). Fig.1(e) shows the electric field distribution of doubly-resonant cavity, where a strong enhancement of mode-localized electric field appears in both the left and right dielectric layers near the metal. Moreover, for the topological protected properties of our doubly-resonant microcavity, the period of Phc1 and Phc2 can be scaled independently. Specifically, as long as we protect the mirror symmetry of Phc1 and Phc2, the topological Tamm states will always exist and their resonant wavelengths can be flexibly tuned with the period change of photonic crystal. Therefore, when we change the period of Phc2 and keep the period of Phc1 unchanged, the shift of the resonant wavelength can be easily tuned by changing the ratio $\eta = \Lambda_2/\Lambda_1$, as shown in Fig.1(g). When $\eta$ changes, there are always two peaks in the transmission spectra, where the position of TTS1 is almost constant whereas TTS2 moves. Specifically, when $\eta > 1$, the $\Lambda_2$ increases and TTS2 is redshifted. Conversely, as $\eta$ reduces, the $\Lambda_2$ decreases and TTS2 is blueshifted. Fig. 1(f) reflects the variation of the wavelengths at resonant modes with the ratio $\eta$, which illustrates the above conclusion more visually.

On this basis, we experimentally verified the enhancement of quantum emission using this doubly-resonant topological Tamm microcavity. We choose rare earths as an example of QEs and place them in the PMMA layer, where is the overlapping region of electric field leading to the enhancement of local electromagnetic field and the quantum emission. We investigate the PL intensity and emission rate of QEs in three different cases, which respectively are doubly-resonant topological Tamm microcavity, single Tamm structure and glass substrate. For these measurements, QEs are excited by a laser with a wavelength of 500 nm with a power of 10 mW. The PL spectra are shown in Fig.2(a), indicating that the PL intensity is largest in the case of the doubly-resonant topological Tamm microcavity and smallest in the case of the glass substrate. And, when in the doubly-resonant topological Tamm microcavity case, it is respectively close to 8 and 3.5 times of the glass substrate and single resonance cases, significantly improving the emission. Besides, the emission decay curves of QEs shown in Fig.2(b) clearly demonstrate that the doubly-resonant topological Tamm microcavity greatly accelerates the emission rate. This conclusion is specifically reflected in the PL lifetime obtained by fitting the decay curve by the double exponential decay function $I(t) = A_1 e^{(-t/\tau_1)} + A_2 e^{(-t/\tau_2)}$. The PL lifetime is defined as $t = \dfrac{A_1 \tau_1^2 + A_2 \tau_2^2}{A_1 \tau_1 + A_2 \tau_2}$. When QEs are in the case of doubly-resonant topological Tamm microcavity, single Tamm structure and glass substrate, the PL lifetime is respectively t1=61.47us, t2=232.15us and t3=502.61us. Based on these results, it can be concluded that when QEs are in the doubly-resonant topological Tamm microcavity case, the emission rate is significantly faster than the other two cases and it is nearly 1 order of magnitude faster than that of the glass substrate case, indicating the great enhancement of the emission.

To explain the enhancement of the quantum emission in such doubly-resonant topological Tamm microcavity, we have exploited a doubly-resonant cavity temporal coupled mode theory. In this theory, TTS1 and TTS2 respectively refer to cavity mode 1 and 2, which are described as a kind of resonance with amplitude $A = \begin{bmatrix} a & b \end{bmatrix}^T$. Among them, the TTS1 corresponds to the higher frequency band, which refers to the pump laser, and TTS2 corresponds to the lower frequency band, which is the emission wavelength of QEs. Due to the quantum emitter with a two-energy system, it enables the coupling of two orthogonal cavity modes with different frequencies. Besides, the internal leap process of QEs is unidirectional, so it can be excited by photons of higher frequencies and converted into photons of lower frequencies, but the inverse condition is not allowed. Therefore, we deem that the coupling of TTS1 and TTS2 cavities is unidirectional, that is, the TTS2 cavity can couple to the TTS1 cavity, while the TTS1 cavity cannot couple in turn. The TTS1 cavity in our structure is driven by an applied pump laser, while the TTS2 cavity is excited by coupling between the two cavities through QEs. And the coupling equations is:

$$\frac{da}{dt} = (j\omega_a - \gamma_1)a + S_+, \quad \frac{db}{dt} = (j\omega_b - \gamma_2)b + \kappa a \tag{1}$$

where $\omega_a$ ($\omega_b$) is the eigen-frequency of TTS1 (TTS2). $S_+$ is the intensity of interaction between TTS1 cavity and the outside environment. $\gamma_{1(2)}$ is the attenuation rate from the radiative damping and can be expressed in terms of the full width at half maximum of resonant wavelength. $\kappa$ represents the coupling coefficient between the two cavities, which is directly proportional to the overlapping integral of electric field: $\kappa = \dfrac{(\varepsilon_0 - \varepsilon_{\text{PMMA}})\omega_0 \int_{\text{PMMA}} d\mathbf{r}^3 \mathbf{E}_1^* \cdot \mathbf{E}_2}{\langle \phi_1/\phi_1 \rangle \langle \phi_2/\phi_2 \rangle}$, $\langle \phi_{1(2)}/\phi_{1(2)} \rangle = \sum_{m \in \text{all}} \int_m d\mathbf{r}^3 \left( \varepsilon_m \mathbf{E}_{1(2)}^* \cdot \mathbf{E}_{1(2)} + \mu_m \mathbf{H}_{1(2)}^* \cdot \mathbf{H}_{1(2)} \right)$, where $\varepsilon_0$ and $\varepsilon_m$ are the absolute permittivity for vacuum and the materials, respectively. $\mu_0$ and $\mu_m$ are respectively the permeability of vacuum and the materials. $E_{1(2)}$ and $H_{1(2)}$ are respectively the electric and magnetic field distribution of resonant modes. In our structure, TTS2 cavity is coupled with the emission wavelength of QEs embedded in. And, the leakage of TTS2 cavity, which are expressed as $|\gamma_2 b|$, an reflect the quantum emission of QEs. To simplify the model, we consider the interaction strength of TTS1 cavity with pump laser as constant 1, then we can obtain the emission amplitude: $\Gamma = \kappa \gamma_2 S_+ / \sqrt{\left[(\omega_l - \omega_a)^2 + \gamma_1^2\right]\left[(\omega - \omega_b)^2 + \gamma_2^2\right]}$, where $\omega_l$ indicates the frequency of the pump laser. Clearly, the quantum emission intensity $\Gamma$ is affected by the pump laser frequency $\omega_l$ and the coupling strength $\kappa$. Fig.3(a) shows the variation of the quantum emission $\Gamma$ with pump laser frequency $\omega_l$, which clearly indicates that when the pump laser wavelength is equal to TTS1 wavelength ($\lambda_1 = 500nm$), the quantum emission intensity greatly increases. The inset in Fig.3(a) shows the relationship between quantum emission intensity at the peak and pump laser wavelength, reflecting the above phenomenon well. Besides, Fig.3(b) shows the variation of the quantum emission $\Gamma$ with the coupling strength $\kappa$. The results clearly demonstrate that the maximum coupling strength exits when the emission wavelength of QEs locates at resonant mode of $\omega_b$ and the pump laser wavelength is at resonant mode of $\omega_a$, leading to a great enhancement of quantum emission.

In experiment, we have measured PL spectrum of QEs under three different cases to verify the above conclusion. One case is that the pump laser wavelength locates at the resonant wavelength, whereas the emission wavelength of QEs is away from the resonant modes. Another case is the opposite condition that the emission wavelength of QEs is at the resonant mode, while the pump laser wavelength is not. In the last case, both the pump laser and emission wavelengths are at the resonant modes of our structure. These three cases were accomplished by two experimental approaches. One of the methods is to remain the doubly-resonant microcavity unchanged but continuously change the wavelength of pump laser with constant power, to observe the variation of quantum emission intensity. The results are shown in Fig.3(c). Another method is to keep the wavelength and power of the pump laser constant, but change Phc2, and results are shown in Fig.3(d). From the experimental PL spectra in Fig.3(c) and Fig.3(d), it clearly exhibits that the PL intensity is maximum when the resonant modes are coupled with the pump laser and emission wavelengths.

Moreover, owing to the asymmetry and the presence of silver film, our doubly-resonant topological Tamm microcavity has the ability of unidirectional emission, which can improve the collective efficiency of quantum emission of the rare earths within it. As shown in Fig.4(a), the transmission spectra of our doubly-resonant topological Tamm microcavity are different in the case of broad-spectrum laser incident from different directions, owing to the asymmetry of our structure. Furthermore, the experimental PL intensity of the rare earths embedded in our structure are shown in Fig.4(b). When the laser is incident from the same direction and the emission of the rare earths is collected from different directions, there is a large difference in PL intensity, which indicates that the emission of the rare earths is obviously unidirectional with the advantage of collecting the emission.

In conclusion, we experimentally realize the tunable topological doubly-resonant cavities to enhance the emission of quantum emitters. When the resonant wavelengths of the doubly-resonant topological Tamm microcavity are coupled with the pump laser wavelength and emission wavelength of quantum emitters, the quantum emission is increased greatly. In particularly, we observe the faster emission rate and the higher efficiency of unidirectional quantum emission, which has promising applications in optical detection, sensing, filtering, and light-emitting devices.

**Funding.** National Key Research and Development Program of China (2017YFA0303702, 2017YFA0205700); National Natural Science Foundation of China (11690033, 61425018, 11621091); Fundamental Research Funds for the Central Universities (14380139).

**Acknowledgments.** We thank for the support from the National Key Research and Development Program of China, the National Natural Science Foundation of China and the Fundamental Research Funds for the Central Universities.

**Disclosures.** The authors declare no conflicts of interest.

**Data availability.** Data underlying the results presented in this paper are not publicly available at this time but may be obtained from the authors upon reasonable request.

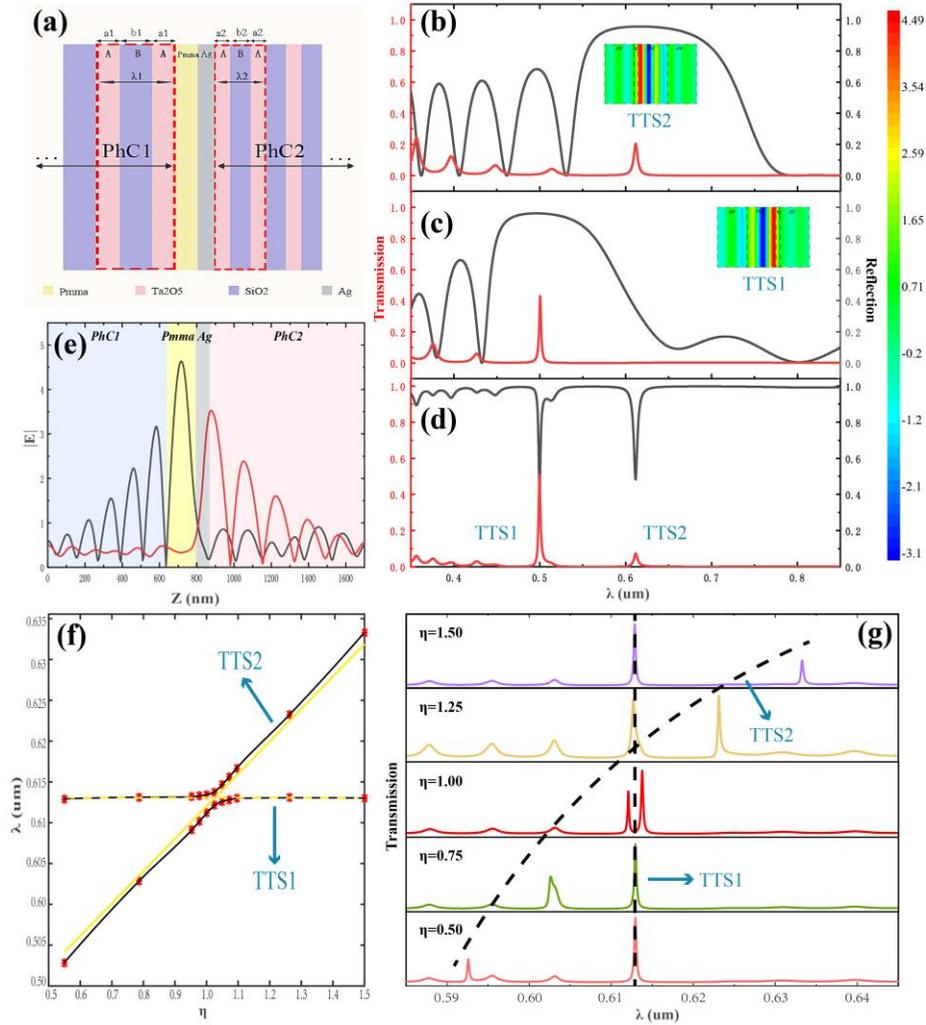

Fig. 1. (a) Sketch of the doubly-resonant microcavity. The period of the Phc1 and Phc2 is $\Lambda_{1(2)}$ with A/B/A as its unit cell as marked by the red dashed lines. Here, layer A is Ta2O5, and layer B is SiO2. Reflection (black) and transmission (red) spectra of the Ag/Phc2 (b), Phc1/Ag (c) and Phc1/PMMA/Ag/Phc2 (d) structures. The sharp peaks represent topological Tamm states. Inset: the distribution of electric field, and an optical localization effect occurs at the interface between metal and dielectric plane. (e) Schematic about the electric field distribution, where the overlapping region of local electromagnetic field enhancement is QEs placed. (f) The variation of resonant wavelengths with ratios $\eta$. (g) Transmission spectra for Phc1/Phc2 structure with different ratios $\eta$.

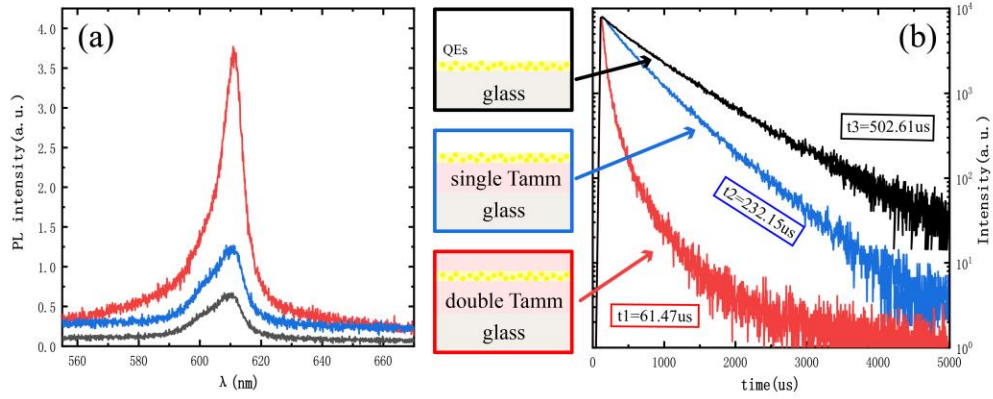

Fig. 2. The experimental data of the PL intensity (a) and the PL lifetime (b) obtained from QEs in three different cases, which respectively are glass substrate (black line), single Tamm (blue line) and double Tamm (red line). In fig. 2(b), the PL lifetime obtained after fitting the decay curve by the double exponential decay function is shown in the legend.

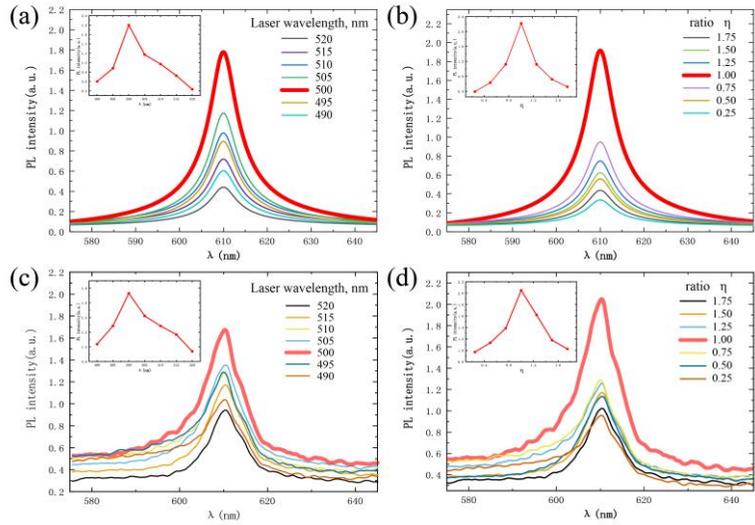

Fig. 3. The variation of PL intensity with $\omega_l$ (a) and $\kappa$ (b). The thickened red line is much higher than the other spectra both in (a) and (b), indicating that the PL intensity is greatly enhanced when resonant modes are coupled with the pump laser wavelength and emission wavelength of QEs. This conclusion is reflected more clearly in the inset. The change of experimental PL intensity with pump laser wavelength (c) and structure parameter (d). This is more clearly in the inset, which is in good agreement with the numerical simulation results of Fig. 3(a, b).

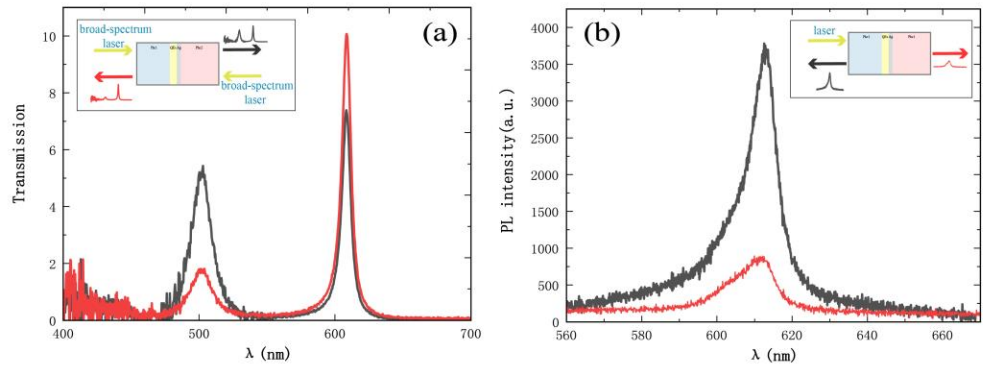

Fig. 4. (a) Experimental transmission spectra in the case of broad-spectrum laser incident from different directions. (b) The PL intensity when the laser is incident from the same direction and the emission of QEs is collected from different directions. The black and red lines represent the position where QEs is collected and the schematics are reflected in the inset.